\documentclass[nofootinbib,reprint,amsmath,amssymb,aps]{revtex4-2}
\usepackage{amsmath,pgf,float,appendix, hyperref,scalerel,amssymb}
\usepackage{footnote,stackengine,tikz,
    graphicx,subcaption,bm,dcolumn,tabularx}
\newcommand{\beq}{\begin{equation}\begin{aligned}}
\newcommand{\eeq}{\end{aligned}\end{equation}}

\begin{document}
\title{Optimization of the Woodcock Particle Tracking Method Using Neural Network}
\author{Bingnan Zhang}%
 \email{zbn@mail.ustc.edu.cn}
\affiliation{University of Science and Technology of China, Hefei, Anhui, China}

\begin{abstract} 
The acceptance rate in Woodcock tracking algorithm is generalized to an arbitrary position-dependent variable $q(x)$. A neural network is used to optimize $q(x)$, and the FOM value is used as the loss function. This idea comes from physics informed neural network(PINN), where a neural network is used to represent the solution of differential equations. Here the neural network $q(x)$ should solve the functional equations that optimize FOM. For a 1d transmission problem with Gaussian absorption cross section, we observe a significant improvement of the FOM value compared to the constant $q$ case and the original Woodcock method. Generalizations of the neural network Woodcock(NNW) method to 3d voxel models are waiting to be explored.
\end{abstract}
\maketitle

\section{Introduction}
\par 
Woodcock particle tracking method was first introduced in neutron transport by Woodcock et al\cite{wood}. Since its invention, many variants of Woodcock method have been implemented in Monte Carlo codes, for example Serpent \cite{serp},MORET\cite{more},MONK\cite{monk},IMPC\cite{impc1}\cite{impc2}. It has also been applied in radioactive transfer\cite{radi}, medical physics \cite{med1}\cite{med2},computer graphics\cite{cgraph1}\cite{cgraph2}\cite{cgraph3}, and other areas. In Monte Carlo simulation, collision distance is usually calculated by inverting the integral of cross section along the particle path. However, this is very difficult when the media is in-homogeneous. Woodcock tracking takes a fictitious cross section $\Sigma_s>=max(\Sigma(x))$ as the majorant to sample collision distance, then the distance is accepted with probability $\Sigma(x)/\Sigma_s$. When the distance is not accepted, we say there is a null collision. Since the sampling cross section $\Sigma_s$ is a constant, it is straightforward to sample the collision distance $d=-ln \zeta/\Sigma_s$, where $\zeta$ follows a uniform distribution on $[0,1]$. However, in problems where there are regions with significantly bigger cross sections than other places, for example the control rods in nuclear reactors, choosing the majorant as the sampling cross section can be cumbersome, as it will generate a lot of unnecessary null collisions. In \cite{leg1}\cite{leg2}\cite{neg}, a biased version of Woodcock tracking is proposed, where the sampling cross section is an arbitrary constant, and the acceptance rate is also a variable. The particle weight is adjusted after every collision to compensate for these changes. This biased version of Woodcock tracking can significantly improve the FOM value, which is defined as $\frac{1}{relative\_error^2\cdot run\_time}$ to measure the efficiency, if the parameters are chosen properly. The acceptance rate can also be a position-dependent function rather than a constant\cite{leg1}. This in principle can further improve the efficiency because there are more degree of freedoms to tune. However, since the problem becomes more complicated, optimization also requires more effort. In this paper, a neural network inspired  by PINN\cite{pinn1}\cite{pinn2} is used to optimize the sampling cross section and the acceptance rate, and they are tested in a 1d transmission problem. 

Section \ref{II} is a general introduction of Woodcock tracking and biased Woodcock tracking.  Section \ref{III} states the transmission problem and analytically derives FOM. Section \ref{IV} and \ref{V} describe the numerical optimization and simulation results. Section \ref{VI} includes a summary and future research directions.

\section{The Woodcock tracking formalism}\label{II}
As described in section I, the Woodcock tracking method is characterized by a majorant cross section $\Sigma_s$, and the distance is sampled from distribution 
\beq
f(x)=\Sigma_se^{-\Sigma_s x}
\eeq
which is accepted with probability $\Sigma(x)/\Sigma_s$. This can be described by the following code:\\
\\
\hspace*{4em} w=1;\\
\hspace*{4em} \textbf{while} !collided:\\
\hspace*{5em} $\zeta_1=rand()$;\\
\hspace*{5em} $s=-ln(\zeta_1)/\Sigma_s$;\\
\hspace*{5em} \textbf{x}=\textbf{x}+\textbf{$\Omega$}s;\\
\hspace*{5em} $\zeta_2=rand()$;\\
\hspace*{5em} \textbf{if} $\zeta_2<\Sigma(\textbf{x})/\Sigma_s$:\\
\hspace*{6em} collided=true;\\
\hspace*{6em} handle real collision;\\
\hspace*{5em} \textbf{else}:\\
\hspace*{6em} null collision,do nothing;\\
\hspace*{5em} \textbf{end}\\
\hspace*{4em} \textbf{end}\\

A mathematical proof of the correctness can be found in \cite{veri}. In biased Woodcock tracking, however, the acceptance rate becomes a variable, which is denoted as $q$. The particle weight is adjusted at every step to compensate for this change: \\
\\
\hspace*{4em} w=1;\\
\hspace*{4em} \textbf{while} !collided:\\
\hspace*{5em} $\zeta_1=rand()$;\\
\hspace*{5em} $s=-ln(\zeta_1)/\Sigma_s$;\\
\hspace*{5em} \textbf{x}=\textbf{x}+\textbf{$\Omega$}s;\\
\hspace*{5em} $\zeta_2=rand()$;\\
\hspace*{5em} \textbf{if} $\zeta_2<q$:\\
\hspace*{6em} collided=true;\\
\hspace*{6em} $w=w\frac{\Sigma(x)}{\Sigma_sq}$;\\
\hspace*{6em} handle real collision;\\
\hspace*{5em} \textbf{else}:\\
\hspace*{6em} $w=w\frac{1-\Sigma(x)/\Sigma_s}{1-q}$;\\
\hspace*{5em} \textbf{end}\\
\hspace*{4em} \textbf{end}\\

The unbiasedness of such a modification can be proved via the probability generating function(PGF) of in-homogeneous Poisson process(IHPP)\cite{leg1}. When $\Sigma_s<\Sigma(x)$, the particle weight turns negative in a null collision event\cite{neg}. These negative weight particles can be understood as fictitious anti-particles that will cancel the over-weight of other particles. Various authors have proposed different forms of $q$\cite{galt}\cite{DPM}, however, little effort has been devoted to its theoretical or numerical optimization. When $q(x)$ is a position-independent constant, a plot of FOM with respect to $\Sigma_s$ and q has been made in \cite{leg2}, and the optimal FOM is about 7 times bigger than Carter's algorithm\cite{cart}. In this paper, we focus on the position-dependent $q(x)$, and tries to optimize it using neural network. This should inspire people to generalize it to 3d voxel models.

\section{The transmission problem}\label{III}
\begin{figure}[h]
\centering
\includegraphics[scale=0.4]{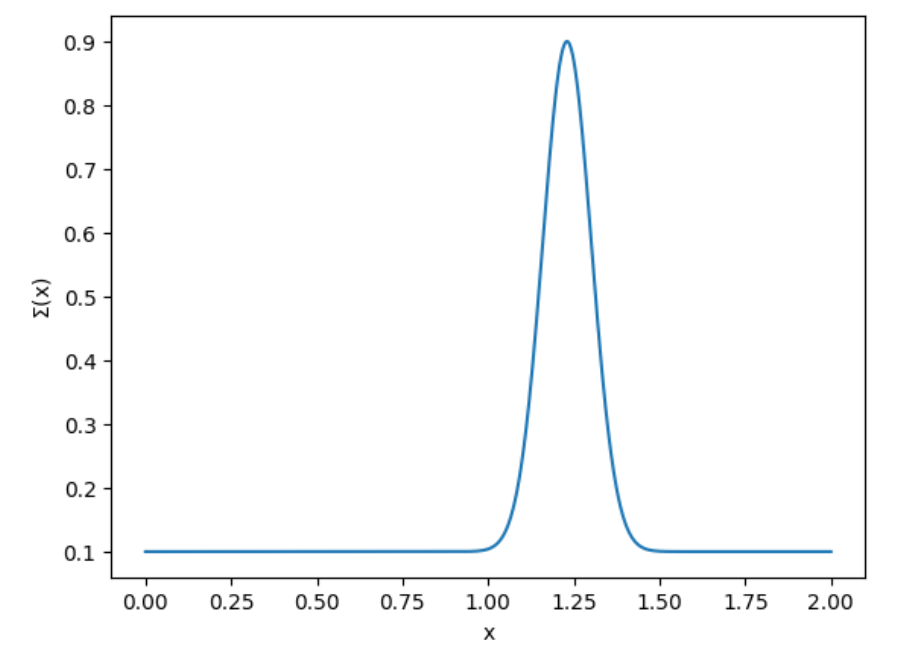}
\caption{The narrow Gaussian cross section $\Sigma(x)$}
\label{fig:sigmax}
\end{figure}

The problem we consider in this paper is a 1d transmission problem with narrow Gaussian absorption cross section  (figure \ref{fig:sigmax}). The absorption peak lies at $x=1.23$.
\beq
\Sigma(x)=0.8e^{-\frac{(x-1.23)^2}{0.1^2}}+0.1
\eeq
Particles fly from $x=0$ to the right and we record the number of particles that are not absorbed within the $0<x<2$ region. The algorithm is summarized below:\\
\\
\hspace*{4em} w=1;\\
\hspace*{4em} \textbf{while} !collided:\\
\hspace*{5em} $\zeta_1=rand()$;\\
\hspace*{5em} $s=-ln(\zeta_1)/\Sigma_s$;\\
\hspace*{5em} x=x+s;\\
\hspace*{5em} \textbf{if} $x>L$:\\
\hspace*{6em} record transmission event;\\
\hspace*{6em} kill the particle;\\
\hspace*{5em} \textbf{else}:\\
\hspace*{6em} $\zeta_2=rand()$;\\
\hspace*{6em} \textbf{if} $\zeta_2<q(x)$:\\
\hspace*{7em} collided=true;\\
\hspace*{7em} kill the particle;\\
\hspace*{6em} \textbf{else}:\\
\hspace*{7em} $w=w\frac{1-\Sigma(x)/\Sigma_s}{1-q(x)}$;\\
\hspace*{6em} \textbf{end}\\
\hspace*{5em} \textbf{end}\\
\hspace*{4em} \textbf{end}\\

Following \cite{leg1}, we derive the relative error of the particle weight $w$.
The expectation value can be expanded with respect to null collision events.
\beq
E[w]=&\sum_{k=0}^{\infty}\prod_{i=1}^{k}\int dx_i E[w|X_1=x_1,X_2=x_2,...,X_k=x_k]\\&
\prod_{i=1}^k(1-q(x_i)) p_{X|k}(x_1,x_2,...,x_k)p_k
\label{ew}
\eeq
where $p_k$ is the probability that the particle undergoes k collisions before reaching $x=2$. $p_{X|k}(x_1,x_2,..,x_k)$ represents the probability that such collisions happen at $x_1,x_2,...,x_k$. The integral doesn't change if we shuffle the order of $x_1,x_2,...,x_k$, so we ignore the constraint that $x_1<x_2<...<x_k$. Since $\Sigma_s$ is constant, the probability density should be uniform\cite{uniform}, so 
\beq
p_{X|k}(x_1,x_2,...,x_k)=1/L^k
\eeq
where $L=2$. $\prod_{i=1}^k(1-q(x_i))$ makes sure that all collisions are null so that the particle is not killed in the middle. $E[w|X_1=x_1,X_2=x2,...,X_k=x_k]$ is the particle weight in such a transmission event. Under the biased Woodcock formalism  \beq E[w|X_1=x_1,X_2=x2,...,X_k=x_k]=\prod_{i=1}^k\frac{1-\Sigma(x_i)/\Sigma_s}{1-q(x_i)}\eeq
Put these into equation \eqref{ew}, we get
\beq
E[w]=\sum_{k=0}^{\infty}(1-\frac{\bar{\Sigma}}{\Sigma_s})^kp_k
\eeq
where $\bar{\Sigma}$ is the mean of $\Sigma(x)$ on the line.
We can also calculate the second moment in the same way
\beq
E[w^2]=&\sum_{k=0}^{\infty}\prod_{i=1}^{k}\int dx_i E[w^2|X_1=x_1,X_2=x_2,...,X_k=x_k]\\&
\prod_{i=1}^k(1-q(x_i)) p_{X|k}(x_1,x_2,...,x_k)p_k
\\=&\sum_{k=0}^{\infty}\prod_{i=1}^{k}\int dx_i 
\prod_{i=1}^{k}\frac{(1-\Sigma(x_i)/\Sigma_s)^2}{1-q(x_i)}\frac{1}{L^k}p_k
\\=&\sum_{k=0}^{\infty}(\frac{1}{L}\int dx \frac{(1-\Sigma(x)/\Sigma_s)^2}{1-q(x)})^kp_k
\eeq
The number of collisions k follows Poission distribution with PGF
\beq
G(z)=\sum_{k=0}^{\infty}z^kp_k=exp((z-1)L\Sigma_s)
\eeq
So the first and second moment of w can be calculated
\beq
E[w]=exp(-\int_0^L dx \Sigma(x))
\eeq
\beq
E[w^2]=exp\{\int_0^L dx \frac{q(x)\Sigma_s-2\Sigma(x)+\Sigma(x)^2/\Sigma_s}{1-q(x)}\}
\eeq
The squared relative error is 
\beq
r^2=&\frac{1}{N}\frac{E[w^2]-E[w]^2}{E[w]^2}
\\=&\frac{1}{N}\left(exp(\int_0^L\frac{q(x)(\Sigma_s-2\Sigma(x))+\Sigma(x)^2/\Sigma_s}{1-q(x)}  )-1 \right)
\label{eq:r2}
\eeq
where N is the number of particles.

The run time is roughly proportional to the number of cross section evaluations \cite{leg1}\cite{leg2}, which is also the number of null collisions. We have 
\beq
T=N\int_0^L dx \Sigma_s(1-q(x)) e^{-\int_0^x\Sigma_s q(y)dy}
\label{eq:t}
\eeq
the exponential $e^{-\int_0^x dy \Sigma_s q(y)}$ is the probability that a particle is not absorbed on $[0,x]$, and $\Sigma_s(1-q(x)) dx$ is the number of null collisions such a particle contributes on $dx$. Note that this expression for $T$ is different from the one used in \cite{leg1}. The derivation of $T$ in \cite{leg1} is attached in appendix \ref{ap}.

Finally, the FOM is defined as
\beq
FOM=&\frac{1}{Tr^2}
\\=&\frac{1}{exp(\int_0^Ldx\frac{q(x)(\Sigma_s-2\Sigma(x))+\Sigma(x)^2/\Sigma_s}{1-q(x)}  )-1}
\\&\cdot\frac{1}{\int_0^L dx \Sigma_s(1-q(x)) e^{-\int_0^x\Sigma_s q(y)dy}}
\label{fom}
\eeq
which measures the efficiency of the algorithm.
One interesting limit is $q=0,\Sigma_s>>\Sigma(x)$. In this limit, null collisions are sampled with high frequency and there are no real collisions. FOM reduces to
\beq
FOM\approx&\frac{1}{exp(\int_0^L\Sigma(x)^2dx/\Sigma_s)-1}\frac{1}{\Sigma_sL}\\
\approx&\frac{1}{L\int_0^L\Sigma(x)^2dx}
\eeq
This limit is not yet covered by the neural network version of $q(x)$. As described in section \ref{IV}, $\Sigma_s$ is obtained from $q(-1)$ that is bounded by the activation function $tanh$. One may change the output layer to cover this limit.

\section{Optimization}\label{IV}
Solving for the optimal $q(x)$ that maximize equation \ref{fom} is an extremely complicated functional optimization problem, and the author doesn't think an analytical solution can be obtained. Instead, numerical methods should be applied. The widely used machine learning packages provide us perfect tools for solving such optimization problems. This is the basic idea behind physics informed neural network(PINN)\cite{pinn1}\cite{pinn2}. In this paper, we use a simple FFN network with $tanh$ activation, whose output range is $[-1,1]$. The input data is position $x$, and the output when $0<x<2$ is interpreted as $2q(x)-1$. The output at $x=-1$ is interpreted as $\Sigma_s/\Sigma_m-1$, where $\Sigma_m$ is the maximum value of $\Sigma(x)$. Obviously, this choice is not unique.

\begin{figure}[h]
\centering
\includegraphics[scale=0.4]{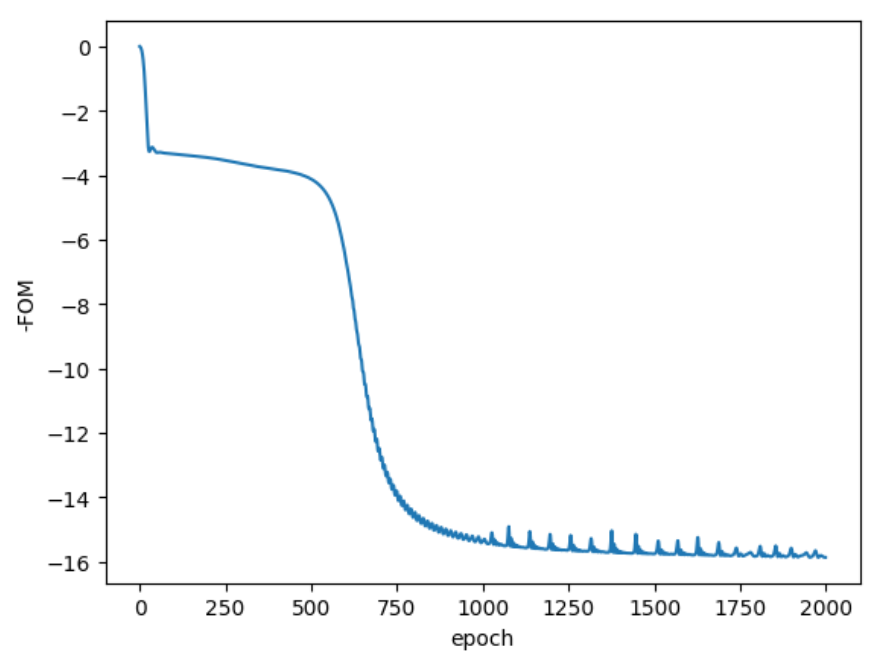}
\caption{The change of FOM with training epochs.}
\label{fig:epoch}
\end{figure}

We discretize the space $[0,2]$ and calculate the integral in equation  \ref{fom} with the Pytorch function trapz. An Adam optimizer is used to maximize FOM. This process is fast because the FFN network used in this paper is small. 
The change of FOM with training epoch is shown in figure \ref{fig:epoch}. 

\begin{figure}[h]
\centering
\includegraphics[scale=0.4]{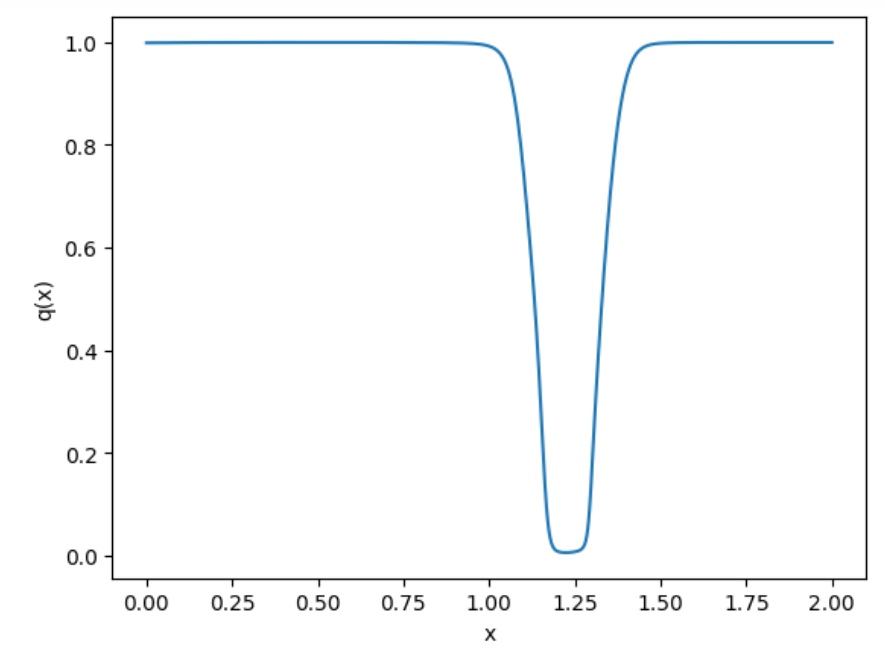}
\caption{q(x) after optimization.}
\label{fig:qx}
\end{figure}

The optimized $q(x)$ is plotted in figure \ref{fig:qx}. Remember in the original version of Woodcock tracking, $q(x)$ is proportional to $\Sigma(x)$ shown in figure \ref{fig:sigmax}. Here the optimized $q(x)$ looks exactly the opposite. $\Sigma_s=0.099$ is obtained from $q(-1)$.

\section{Simulation results}\label{V}
\begin{figure}[h]
\centering
\includegraphics[scale=0.4]{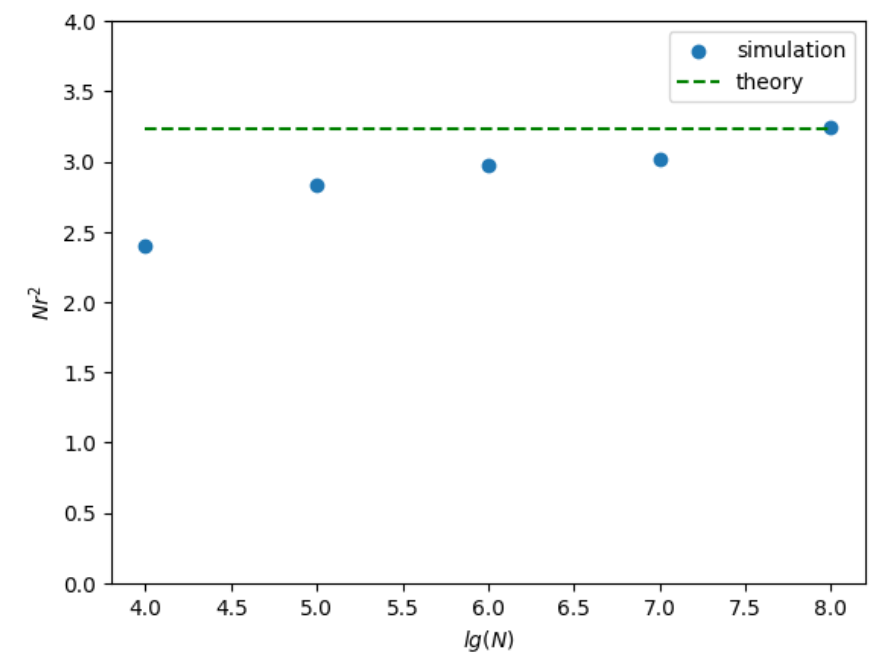}
\caption{The convergence of $Nr^2$ towards theoretical prediction.}
\label{fig:r2}
\end{figure}

\begin{figure}[h]
\centering
\includegraphics[scale=0.4]{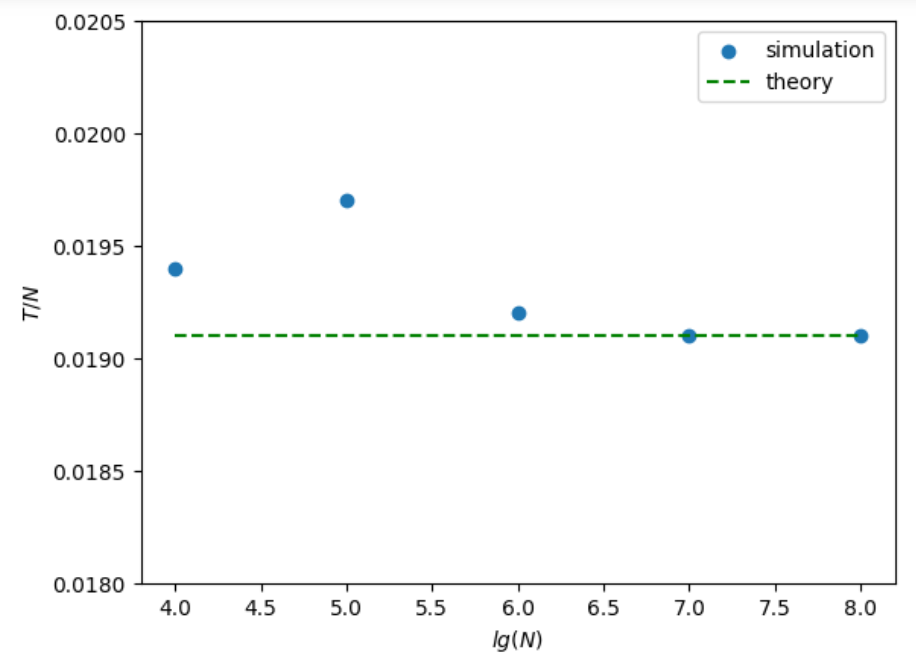}
\caption{The convergence of $T/N$ towards theoretical prediction.}
\label{fig:t}
\end{figure}

Simulations are conducted using several different Woodcock-type tracking methods. The neural network $q(x)$ obtained in section \ref{IV} is used to run the neural network Woodcock(NNW) simulation. The relative error $r$ of transmission rate and the number of cross section evaluations $T$ are recorded. The change of $Nr^2$ and $T/N$ with particle number $N$, where $N$ is introduced to normalize the result, are plotted in figure \ref{fig:r2},\ref{fig:t}. The theoretical predictions are calculated from equation \ref{eq:r2},\ref{eq:t}. The relative error and cross section evaluation number converge to the theoretical values when $N>10^7$, and the correctness of the derivations in section \ref{III} are proven.

 A summary of $Nr^2$, $T/N$, $FOM$ for different Woodcock-type tracking methods is made in table \ref{tab:cmp1}.  

\begin{table}[h]
\centering
\caption{Comparison of the original Woodcock, constant q  biased Woodcock(CBW), and neural network Woodcock(NNW) methods using narrow Gaussian cross section.}
\label{tab:cmp1}
\begin{tabular}{cccc}
\hline\hline\noalign{\smallskip}
\ &Woodcock&CBW&NNW\\
\noalign{\smallskip}\hline\noalign{\smallskip}
$Nr^2$&0.104&0.000727&2.94\\
\noalign{\smallskip}\hline\\
T/N&1.27&89.8&0.0198\\
\noalign{\smallskip}\hline\\
FOM&1.95&3.91&17.2\\
\noalign{\smallskip}\hline\\
\end{tabular}
\end{table}

The first column in table \ref{tab:cmp1} represents the original Woodcock, the second column is the biased Woodcock using constant $q,\Sigma_s$, the third column is the biased Woodcock using the optimized neural network $q(x)$. The constant biased Woodcock method is also used in \cite{bel}, but the parameters are not optimized. Here a grid search is conducted to find the optimal $q,\Sigma_s$, and the second column records the best result. 

The constant biased Woodcock is about twice as efficient as the original Woodcock, while using function $q(x)$ and neural network further improves the efficiency by $340\%$. Note that the efficiency improvement comes from $T$ while the relative error gets bigger.
\begin{table}
\centering
\caption{Comparison of constant q biased Woodcock(CBW) and neural network Woodcock(NNW) using different cross sections. FOM values are recorded.}
\label{tab:cmp2}
\begin{tabular}{cccc}
\hline\hline\noalign{\smallskip}
$\Sigma(x)$&CBW&NNW&Improvement\\
\noalign{\smallskip}\hline\noalign{\smallskip}
$0.8e^{-\frac{(x-1.23)^2}{0.1^2}}+0.1$&3.91&17.2&340\%\\
\noalign{\smallskip}\hline\\
$0.8e^{-\frac{(x-1.23)^2}{1^2}}+0.1$&1.40&1.91&36.4\%\\
\noalign{\smallskip}\hline\\
$\sin(3x)+1$&0.151&0.408&170\%\\
\noalign{\smallskip}\hline\\
$e^{-x}$&1.24&1.84&48.4\%\\
\noalign{\smallskip}\hline\\
\end{tabular}
\end{table}

In table \ref{tab:cmp2}, more simulations are done using several different cross sections and the FOM values are recorded. The optimal constants are already used in constant biased Woodcock. The neural network Woodcock beats constant biased Woodcock in all cases, and the FOM improvement gets bigger when the cross section peaks become sharper. 

Sometimes the constant q biased Woodcock method gives the largest FOM in the $q=0,\Sigma_s>>1$ limit. Obviously, this limit is not covered by the neural network because the range of $\Sigma_s$ is constrained by the $tanh$ activation function. Using other functions that covers the whole $\Sigma_s>0$ region in the output layer should in principle further improve the efficiency.

\section{Conclusion}\label{VI}
In this paper, we optimize the acceptance probability $q(x)$ and sampling cross section $\Sigma_s$ in biased Woodcock tracking method. A neural network is used to represent these variables, and the FOM value is used as the loss function. In a 1d transmission problem with narrow Gaussian absorption cross section, we observe a significant improvement of the FOM value compared to the constant $q$ biased Woodcock and also the original version of Woodcock tracking. This improvement mainly comes from the significant reduction of cross section evaluations. The correctness of the theoretical derivations using IHPP process is supported by the simulation result. Several other cross sections are also used to run the simulation and the efficiency improvement is demonstrated. One may want to change the output layer to cover the $\Sigma_s>>\Sigma(x)$ limit and other optimization methods without neural network can also be explored. In order to make this method more applicable in industrial simulations, future research can also be done on generalizing this method to 3d voxel models. In this case, it might be difficult to obtain the analytical expression of FOM, and a self-adaptive method could be more realistic.

\acknowledgements
The author would like to thank Dr.Sida Gao for his references.

\appendix
\section{}\label{ap}
A different type of computational cost estimation formula is given in \cite{leg1}. This appendix includes their derivation. The computational cost mainly comes from the effort to generate a real collision, so we estimate the number of null collisions generated before a real collision is accepted. Let C be the number of null collisions until a real collision is generated. The computational cost is estimated by the expectation of C as follows:
\beq
E[C]=\int_0^{\infty}\left(\int_0^x\Sigma_sdy\right) q(x)\Sigma_s exp\left(-\int_0^xq(y)\Sigma_sdy\right)dx
\eeq
This works when the space is infinite. When the space is finite, however, the particle can escape the region before a real collision happens. A modification of the above equation gives
\beq
E[C_L]=&\int_0^L\left(\int_0^x\Sigma_sdy\right)q(x)\Sigma_s exp\left(-\int_0^xq(y)\Sigma_sdy\right)dx\\&+\int_0^L\Sigma_sdx\cdot exp\left(-\int_0^Lq(x)\Sigma_sdx\right)
\eeq
When $\Sigma_s$ is a constant, the above equation reduces to
\beq
E[C_L]=&\int_0^L\Sigma_s^2x q(x) exp\left(-\int_0^xq(y)\Sigma_sdy\right)dx\\&+\Sigma_sL\cdot exp\left(-\int_0^Lq(x)\Sigma_sdx\right)
\eeq
where $L$ is the length of the segment.


\begin{thebibliography}{99}
\bibitem{leg1}B.Molnar,D.Legrady, {\it Variance analysis of Woodcock type tracking},International Conference on Mathematics and Computational Methods Applied to Nuclear Science and Engineering(2017)
\bibitem{leg2}B.Molnar, G.Tolnai, and D.Legrady, {\it Variance reduction and optimization strategies in a biased
Woodcock particle tracking framework}, Nuclear Science and Engineering, 190, 56–72(2018).
\bibitem{neg}D. Legrady,B. Molnar,M. Klausz and T. Major,{\it Woodcock tracking with arbitrary sampling cross section using negative weights},Annals of Nuclear Energy,102,116-123(2017).
\bibitem{bel}H.Belangera, D.Mancusib, A.Zoiac,{\it Review of Monte Carlo methods for particle transport in
continuously-varying media}, Eur. Phys. J. Plus 135:877(2020) 
\bibitem{wood}E. Woodcock et al., {\it Techniques used in the GEM code for Monte Carlo neutronics calculations in reactors and other systems of complex geometry}, Proc. Conf. Appl. Comput. Methods Reactor Probl., 557, 2 (1965);
\bibitem{cart}L.Carter, E.Cashwell, and W.Taylor, {\it Monte Carlo sampling with continuously varying cross section along flight paths}, Nuclear Science and Engineering, 48,4, 403–411 (1972).
\bibitem{veri}V.S.Antyufeev, {\it Mathematical verification of the Monte Carlo maximum cross-section technique}, Monte Carlo Methods Appl., 21, 4, 275 (2015)
\bibitem{uniform}P.V.Mieghem,{\it Performance analysis of complex networks and systems},Cambridge University Press(2009).
\bibitem{serp}J.Leppanen, {\it Performance of Woodcock delta-tracking in lattice physics applications using the Serpent Monte Carlo reactor physics burnup calculation code}, Ann.
Nucl. Energy, 37, 5, 715 (2010);
\bibitem{more}B.Forestier et al.,{\it Criticality calculations on Pebble Bed HTR-PROTEUS configuration as a validation for the psedo-scattering tracking method implemented in the MORET 5 Monte Carlo Code}, Int. Conf. Phys. Reactors,Interlaken, Switzerland, September 14–19, 2008.
\bibitem{monk}S.D.Richards et al., {\it MONK and MCBEND: current
status and recent developments}, Ann. Nucl. Energy, 82,
63 (2015);
\bibitem{impc1}P.Fang,X.Wu,Y.Yang,H.Wang,L.Yang,Y.Guo and H.Lai {\it Development and preliminary verification of a Monte Carlo neutron transport code IMPC-Neutron},Annals of Nuclear Energy,175,109221(2022)
\bibitem{impc2}P.Fang,Y.Yang,H.Wang,L.Yang,Y.Guo,H.Lai,and X.Wu,{\it Implementation and optimization of WDT algorithm on IMPC-Neutron and comparison of efficiency of various algorithms},Annals of Nuclear Energy,187,109786(2023)
\bibitem{cgraph1}J.Novak, A.Selle, and W.Jarosz, {\it Residual ratio
tracking for estimating attenuation in participating media},
ACM Transactions on Graphics (TOG), 33, 6, 179 (2014).
\bibitem{cgraph2} M.Raab, D.Seibert, and A.Keller, {\it Unbiased
global illumination with participating media},Monte
Carlo and Quasi-Monte Carlo Methods 2006, Springer,
pp. 591–605 (2008).
\bibitem{cgraph3}R.Sosan,M.M.Movania and S.Siddiqui,{\it Perceptual analysis of distance sampling and transmittance estimation techniques in biomedical volume visualization},Turkish J. Electr. Eng. Comput. Sci.,30,2109-2123(2022)
\bibitem{radi} V.Eymet,D.Poitou,M.Galtier,M.Elhafi,G. Terree, and R.Fournier, {\it Null-collision meshless Monte-Carlo-Application to the validation of fast radiative transfer solvers embedded in combustion simulators}, Journal of Quantitative Spectroscopy and Radiative Transfer, 129, 145–157 (2013).
\bibitem{med1} A. BADAL and A. BADANO, {\it Monte Carlo simulation of
X-ray imaging using a graphics processing unit}, 2009
IEEE Nuclear Science Symposium Conference Record
(NSS/MIC), IEEE (2009), pp. 4081–4084.
\bibitem{med2}Z.Liu, C.Zheng, N.Zhao, Y.Huang, J.Chen, Y.Yang, {\it A GPU-accelerated Monte Carlo dose computation engine for small animal radiotherapy},Medical Physics,50,5238-5247(2023)
\bibitem{wdt}L.WlG.Morgan and D.Kotlyer,{\it Weighted Delta Tracking for Monte Carlo Particle Transport}, Ann.Nucl.Energy,85,1184(2015) 
\bibitem{galt}M.Galtier et al.,{\it Integral Formulation of Null Collision Monte Carlo Algorithms},J. Quant. Spectrosc.Radiat. Transf., 125, 57 (2013); 
\bibitem{DPM}L. SZIRMAY-KALOS et al., {\it Unbiased Light Transport Estimators for Inhomogeneous Participating Media},MOLNAR et al. Comput. Graphics Forum, 36, 2, 9 Wiley Online Library
(2017)
\bibitem{pinn1}H.Hu,L.Qi and X.Chao,{\it Physics-informed Neural Networks (PINN) for computational solid mechanics: Numerical frameworks and applications},Thin-Walled Structures,205,112495(2024)
\bibitem{pinn2}G.E.Karniadakis,Y.Kevrekidis,L.Lu,P.Perdikaris,{\it Physics-informed machines learning}. Nature Reviews Physics(2021)
\end{thebibliography}
\end{document}